\documentclass[prb,preprint]{revtex4-1} 

\usepackage{amsmath}  
\usepackage{amsfonts} 
\usepackage{graphicx} 

\begin{document}

\title{Shortcut to stationary regimes: a simple experimental demonstration}

\author{S. Faure}
\affiliation{Laboratoire de Collisions Agr\'egats R\'eactivit\'e, CNRS UMR 5589, IRSAMC, Universit\'e Paul Sabatier, 118 Route de Narbonne, 31062 Toulouse CEDEX 4, France}

\author{S. Ciliberto}
\affiliation{Universit\'e de Lyon, CNRS, Laboratoire de Physique de l'\'Ecole Normale Sup\'erieure, UMR5672, 46 All\'ee d'Italie, 69364 Lyon, France}

\author{E. Trizac}
\affiliation{LPTMS, CNRS, Univ. Paris Sud, Universit\'e Paris-Saclay, 91405 Orsay, France}

\author{D. Gu\'ery-Odelin}
\affiliation{Laboratoire de Collisions Agr\'egats R\'eactivit\'e, Universit\'e Paul Sabatier,  118 Route de Narbonne, 31062 Toulouse CEDEX 4, France}

\date{\today}

\begin{abstract}
We introduce a reverse engineering approach to drive a RC circuit. This technique is implemented experimentally 1) to reach a stationary regime associated to a sinusoidal driving in very short amount of time, 2) to ensure a fast discharge of the capacitor, and 3) to guarantee a fast change of stationary regime associated to different driving frequencies.  This work can be used as a simple experimental project dedicated to the computer control of a voltage source. Besides the specific example addressed here, the proposed method provides an original use of simple linear differential equation to control the dynamical quantities of a physical system, and has therefore a certain pedagogical value. 
\end{abstract}

\maketitle

In most basic textbooks in electricity, the use of time-dependent voltage source to drive a circuit is reduced to a sinusoidal driving
\cite{purcell}. This is of paramount importance to introduce the concept of filtering in Fourier space \cite{rlc}, a technique that appears in many other fields of physics \cite{thermal} including wave optics \cite{optics}. Such time-dependent circuits also provide an opportunity to train the students on solving linear differential equations, and gives the opportunity to discuss the mechanical equivalent of an inductance, a capacitor or a resistor.

In this article, we propose to revisit the standard RC series circuit subjected to a sinusoidal driving in order to present a reverse use of the differential equation that governs the time evolution of the capacitor charge. More precisely, we show explicitly how the proper shaping of the voltage enables one to reach the stationary regime associated to a sinusoidal driving in a time much shorter than the characteristic time of the circuit. Similarly, we explain how this technique can be extended to the fast discharge of a capacitor or to the sudden change of the driving frequency. 
Here, fast refers to a time scale, chosen {\it a priori}, that can be arbitrarily small. We detail the experimental implementation of those ideas that are well adapted to experimental classes involving a computer control of an instrument, a voltage source in this case. 

The method is generic and is directly inspired by the reverse engineering technique developed in the growing field of Shortcuts To Adiabaticity \cite{RevueSTA,PRX} with applications in classical mechanics \cite{Jarzynski,PRA2014,PRA2016NL}, optical devices \cite{Lin2012}, quantum \cite{Couvert2008,Schaff2011,Walther2012,Bowler2012,Bason2012,Rohringer2015,Du2016,Zhou2017} and statistical physics \cite{Boltzmann,NaturePhys}.

\begin{figure}[t!]
    \begin{center}
        \includegraphics[width=10cm]{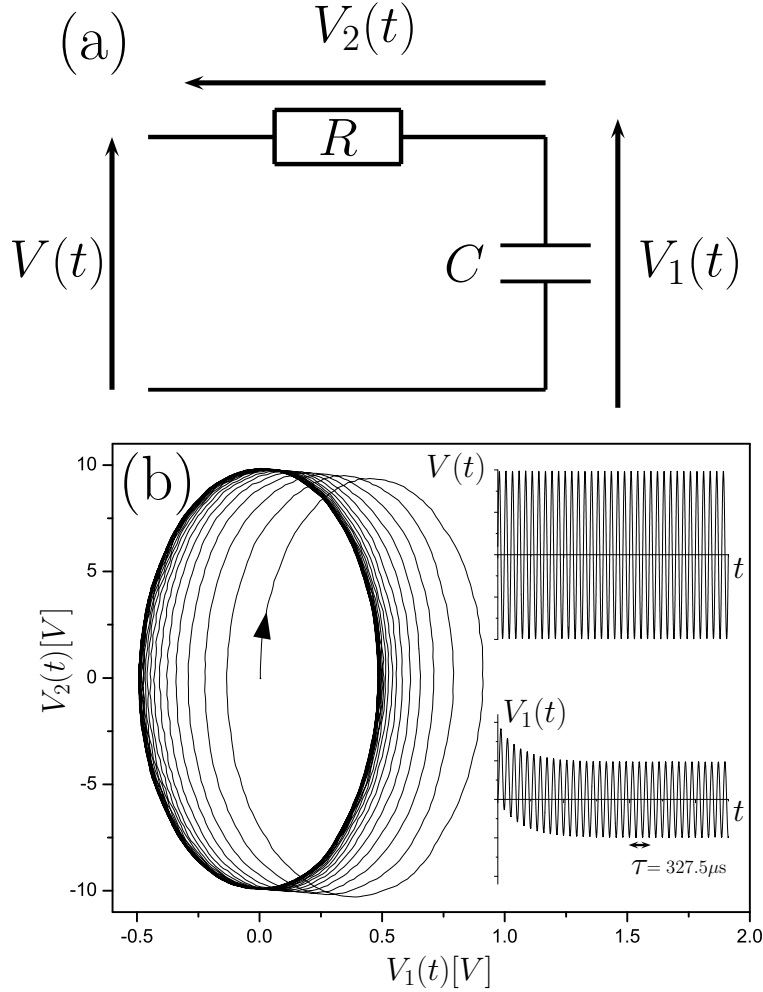}
    \end{center}
    \caption{(a) The RC circuit under study. (b) Phase portrait of our system, displaying the measured evolution of the voltage $V_2(t)=\tau \dot V_1(t)$ as a function of $V_1(t)$ for a voltage source driving $V(t)=V_0 \sin (\omega t)$. Here, $V_0=10$ V, $\omega/2\pi=10$ kHz and 10 000 experimental data points have been gathered to produce the curve, with subsequent
    average over 20 realizations. The insets provide the explicit variation with time of the voltage source, $V(t)$, and the voltage drop across the capacitor, $V_1(t)$.}
    \label{fig1}
\end{figure}

We consider the most simple electric circuit made of a resistor placed in series with a capacitor \cite{RC} driven by a time dependent voltage 
source (see Fig.~\ref{fig1}a). The charge obeys the first order differential equation:
\begin{equation}
\dot q(t) + \frac{q(t)}{\tau} = \frac{V(t)}{R}
\label{eqrc}
\end{equation}
with $\tau=RC$. 
For a sinusoidal driving, 
\begin{equation}
V(t)=V_0 \sin (\omega t),
\label{forcedvoltage}
\end{equation}
 the solution of Eq.~(\ref{eqrc}) is given by the superposition of the response with the source $V$ set  to zero and the forced response: $q(t)=q_{0}(t) + q_f(t)$. In mathematical language, we call these two responses the homogeneous and the particular solutions. The homogeneous solution reads $q_0(t)=A_0 \exp (-t/\tau)$, while the particular solution can be searched in the form 
\begin{equation}
  q_f(t)=A_1^\omega \sin(\omega t)+A_2^\omega \cos(\omega t).
\label{forcedcharge}
\end{equation}
   We readily find $A_1^\omega=(V_0\tau/R)/(1+ \omega^2\tau^2)$ and $A_2^\omega=-\omega \tau A_1$. Using the amplitude phase notation $q_f(t)=A \sin (\omega t - \varphi)$ with $\varphi={\rm ArcTan}(\omega \tau)$ and $A=(V_0\tau/R)/\sqrt{1+ \omega^2\tau^2}$.  With this notation, we clearly see the existence of a time delay, $\varphi/\omega$, between the driving and the response obtained through the time evolution of the charge. It is worth noticing that the forced solution is a particular solution of the second order differential equation without dissipation:
\begin{equation}
 \ddot q_f + \omega^2 q_f=0.  
\label{forcedeq}
\end{equation}

   The constant $A_0$ is determined by the initial condition on the full solution. Assuming that the charge is zero initially, $q(0)=0$, we find :
\begin{equation}
q(t) \,= \, \frac{V_0\tau/R}{1+ \omega^2\tau^2} \,\left\{\, \sin(\omega t)-\omega \tau \big[\cos(\omega t)  - e^{-t/\tau} \,\big] \right\}.
\end{equation}
The transient regime lasts over the time interval for which the term $ e^{-t/\tau}$ is not negligible with respect to one. The time required to reach the stationary regime is thus about six times the characteristic time $\tau=RC$ of the circuit. 
In the limit $\omega \tau \gg 1$, the charge, and therefore the current, undergo a large number of oscillations before reaching the stationary regime. 
This transient towards the stationary regime is most conveniently observed in the so-called phase space ($q$, $\dot q$) \cite{phaseportrait}: the system converges towards an elliptical attractor whose size is dictated  by the driving and the characteristics of the circuit:
\begin{equation}
\left(q^2(t) + \frac{1}{\omega^2}\left(\frac{dq}{dt}\right)^2 \right)\underset{t \gg \tau}{\longrightarrow}   \frac{(V_0\tau/R)^2}{1+ \omega^2\tau^2}. 
\end{equation}

In Figure \ref{fig1}b, we have reconstructed such a phase space by plotting the voltage $V_2(t)$ proportional to $dq/dt$ as a function of $V_1(t)$ proportional to $q(t)$ for the following experimental parameters: $V_0=10$ V, $R=9.9863 \times10^3 \pm 1.1$ $\Omega$, $C=32.8\pm 0.49$ nF (measured with a multimeter Agilent 34405A), $\tau=RC=327.5$ $\mu$s, $\omega = 2\pi \times 10000$ Hz and an acquisition time of $35\times 2\pi/\omega$. 

The dimensionless parameter $\omega \tau = 20.5$ has been chosen sufficiently large to ensure that the system 
undergoes a significant number of oscillations before reaching the stationary regime. The voltage $V_1(t)$ has been recorded using a LeCroy Wavesurfer44Xs oscilloscope (10000 data points are acquired) and averaged over 20 repetitions of the protocol. The voltage $V_2(t)$ cannot be obtained directly since both the voltage source and the oscilloscope that reads the $V_1(t)$ voltage are connected to the ground. We therefore inferred the voltage $V_2(t)$ by performing numerically the substraction: $V_2(t)=V(t)-V_1(t)$. We observe also on the phase plot the well-known clockwise rotation of the trajectory together with the limit cycle (visible as the thick ellipse), that sets in at long time.

\begin{figure}[t!]
    \begin{center}
        \includegraphics[width=10cm]{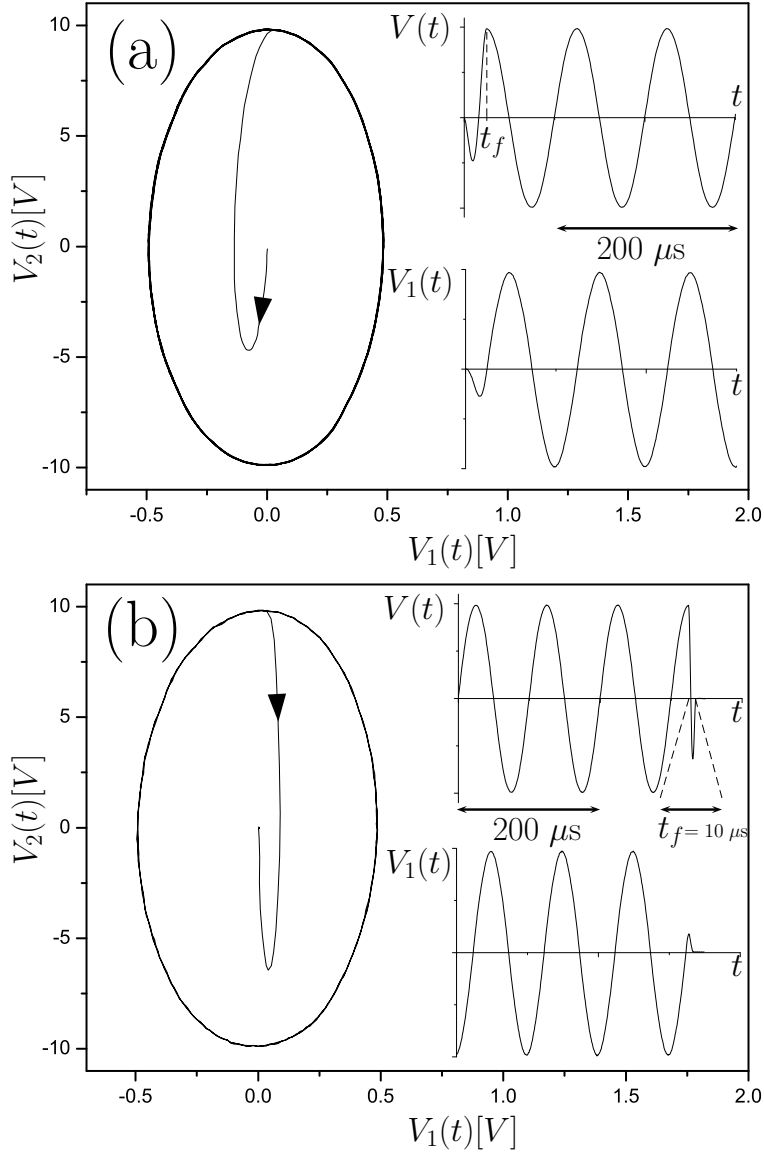}
    \end{center}
    \caption{(a) Experimental evolution of the voltage $V_2(t)=\tau \dot V_1(t)$ as a function of $V_1(t)$ for a shaped voltage $V(t)$ imposed to the RC circuit. Here,
    the target time $t_f$ is chosen to be $t_f=\pi/(2 \omega) = 25 \,\mu$s.
    From $t$ equal to zero to $\pi/(2 \omega)$, the signal $V(t)$ has been calculated to force the evolution of the charge towards the stationary regime, and to be continuously connected to the sinusoidal driving voltage for $t\geq t_f$. (b) Experimental evolution of the voltage $V_2(t)$ as a function of $V_1(t)$ for the driving $V(t)$ which ensures the discharge of the capacitor in $t_f=$10 $\mu$s for an initial charge in the stationary regime associated to the driving frequency 10 kHz. The amplitude and frequency are the same as for Fig.~\ref{fig1}. Insets represent the voltages $V(t)$ and $V_1(t)$ as a function of time.}
    \label{fig2}
\end{figure}

In the following, we propose to engineer the voltage source to reach the stationary regime on a much shorter amount of time $t_f\ll \tau$. For $t>t_f$, the voltage will be the driving voltage (\ref{forcedvoltage}). Within our approach, $t_f$ can be chosen at will, in principle arbitrary small. We adopt a reverse engineering approach. To this end, we first fix the boundary conditions that we would  like on the charge $q(t)$: $q(0)=0$, and $q(t_f)=q_f(t_f)$, $\dot q(t_f)=\dot q_f(t_f)$ and $\ddot q(t_f)=\ddot q_f(t_f)=-\omega^2 q_f(t_f)$. The last condition is important since the stationary trajectory we aim at reaching is solution of the second order linear differential equation (\ref{forcedeq}). We add the two following constraints $\dot q(0)=0$, $\ddot q(0)=0$ to ensure a smooth initial variation of the charge. 
As the motion of the charge is sinusoidal, the boundary conditions on the first and second derivative should be chosen consistently. 
The second step consists in choosing an interpolation function for the charge. In practice and for the sake of simplicity, we take a fifth order polynomial:
\begin{eqnarray}
q(t)  & = & \left[  10q(t_f) - 4t_f \dot q(t_f) + t_f^2 \ddot q(t_f)/2   \right] \left( \frac{t}{t_f}\right)^3 \nonumber \\
& + & \left[   -15q(t_f) + 7t_f \dot q(t_f) - t_f^2 \ddot q(t_f)  \right] \left( \frac{t}{t_f}\right)^4\nonumber \\
& + & \left[   6q(t_f) - 3t_f \dot q(t_f) + t_f^2 \ddot q(t_f)/2   \right] \left( \frac{t}{t_f}\right)^5. \label{eqq}
\end{eqnarray}
By plugging this time-dependent form for the charge into the equation (\ref{eqrc}), we infer the voltage $V(t)$ that we should impose to the circuit
to obtain the desired evolution of the charge. This is the essence of the reverse engineering technique.

As a concrete example, we propose to reach the stationary regime in a quarter of the driving period $t_f=\pi/(2 \omega)$ (see Fig.~\ref{fig2}a). As a result, we fix the final values for the charge $q(t_f)=A_1^\omega$, $\dot q(t_f)=-\omega A_2^\omega$, $\ddot q(t_f)=-\omega^2 q(t_f)$. With such boundary conditions, we have found the following voltage for the time interval $0 \leq t \leq t_f$:
\begin{eqnarray}
V(t) & = & -\frac{V_0}{2} \left( \frac{t}{t_f} \right)^2\frac{1}{1+\omega^2\tau^2} \bigg\{ a_2 +a_3 \left( \frac{t}{t_f} \right)  \nonumber \\
&+&  a_4 \left( \frac{t}{t_f} \right)^2 + a_5  \left( \frac{t}{t_f} \right)^3   \bigg\}
\end{eqnarray}
with
\begin{eqnarray}
a_2 & = & 3 \left( \frac{\tau}{t_f} \right) \left[  -20 + 8 \omega^2 \tau t_f + (\omega t_f)^2\right], \nonumber \\
a_3 & = & 120\frac{\tau}{t_f} +  (\omega t_f)^2 -4(5+14(\omega \tau)^2),\nonumber \\
a_4 & = &-60\frac{\tau}{t_f} -2 (\omega t_f)^2-9 \omega^2 \tau t_f +30(1+(\omega \tau)^2),\nonumber \\
a_5 & = & -12 + 6 \omega^2 \tau t_f  + (\omega \tau)^2,
\end{eqnarray}
that meets all our boundary conditions requirements. For $t\geq t_f$, the voltage is simply the sinusoid of the voltage driving (see Eq.~(\ref{forcedvoltage})). To drive the voltage source, $V(t)$, with such a time dependency, we use the LabVIEW control of the arbitrary waveform generator Keysight 33611A (see upper insets of Figs.~\ref{fig2}a). The imposed source voltage is discretized with a time step of 2.5 ns. Interestingly, our fast protocol for the chosen boundary values does not exhibit a voltage overshoot: the designed voltage has an amplitude always smaller or equal to $V_0=10$ V in our experiment.
The resulting measured voltages are summarized in the phase space plot of Figure \ref{fig2}a. As expected we observe the rapid convergence towards the stationary regime. Comparing the inset of Figs.~\ref{fig2}a, one clearly sees that the charge time evolution measured through $V_1(t)$ responds to the change of the voltage source $V(t)$ with a delay. It is worth noticing that the convergence towards the stationary regime has been dramatically accelerated thanks to our protocol as it can be visually observed by comparing Figs.~\ref{fig1}b and \ref{fig2}a. The stationary regime is approximately reached in a time $6\tau \simeq 2$ ms when the voltage source is applied suddenly while our fast protocol requires a quarter of a period $\pi/2\omega =25 \; \mu$s. The gain in time is therefore about two orders of magnitude. We have taken boundary conditions at final time for a quarter of period just for convenience and simplicity, but the method still holds for shorter amount of time. 

In a similar manner, the reverse transformation from the stationary regime to the complete discharge of the capacitor can also be driven on a short amount of time. The calculation of the desired voltage is obtained in the very same manner using the proper boundary conditions for the charge. Figure~\ref{fig2}b (lower panel) illustrates such an experimental realization with the same electrical circuit using $V_0=10$ V, $\omega/2\pi=10$ kHz and $t_f=10$ $\mu$s. 

\begin{figure}[t!]
    \begin{center}
        \includegraphics[width=10cm]{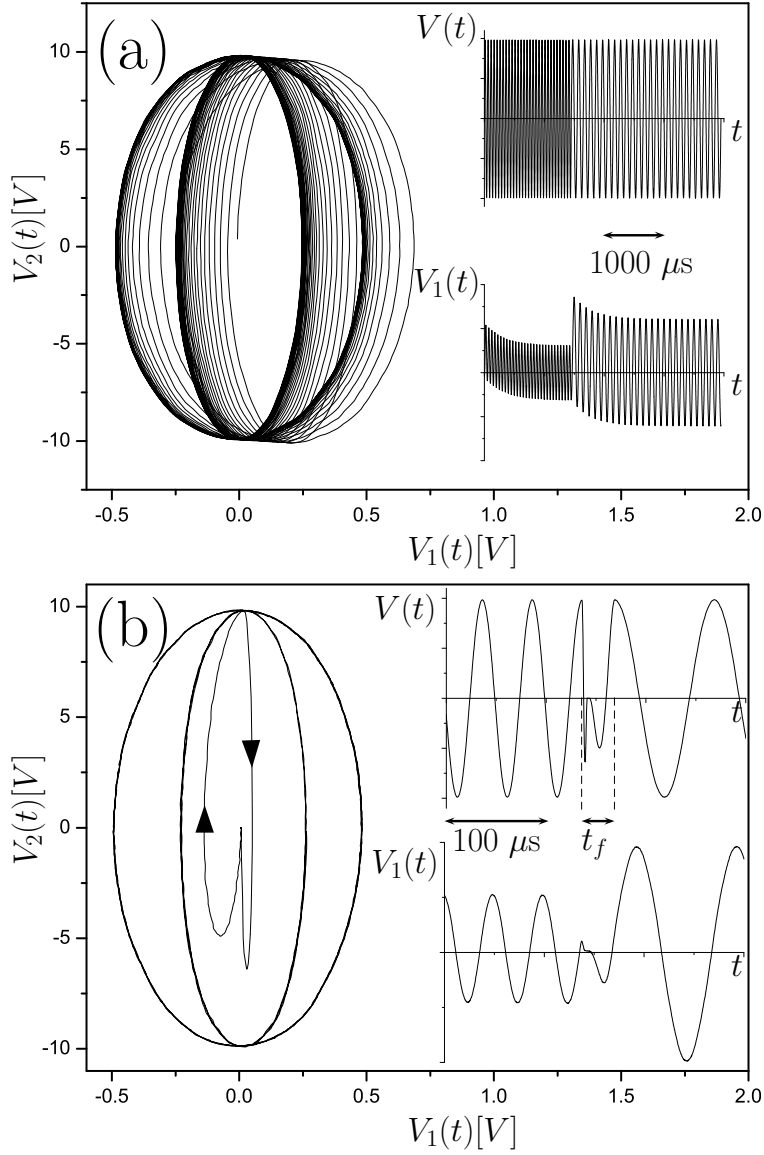}
    \end{center}
    \caption{(a) Phase portrait representation ($V_2(t)$ as a function of $V_1(t)$). The circuit undergoes a sudden frequency change from a sinusoidal driving frequency at 20 kHz to 10 kHz. 
    We observe the convergence towards the initial stationary state (internal ellipse) to the targeted one (external ellipse). (b) Similar plot using the reverse engineering technique 
    to accelerate the change of stationary regime, and operate the switch in a chosen time $t_f$. 
    The voltage is engineered in a non sinusoidal manner during a time span $t_f=\,10+25=\,$35 $\mu$s to ensure first the passage from 
    the stationary regime at 20 kHz to a complete discharge, and then from $q=0$ to the stationary regime associated to the frequency 10 kHz. Same notation as in Fig. \ref{fig2} for the insets.}
    \label{fig3}
\end{figure}

Combining the previous methods, one can readily extend the control of the circuit driving to connect two stationary states associated to two different frequency driving,
going through the state of ``rest'' (vanishing $V_1$ and $V_2$) as an intermediate. We have realized this experiment by driving the system at 20 kHz and then at 10 kHz as explicitly shown in Fig.~\ref{fig3}. We present in the upper panel such a transformation performed with a sudden change of the frequency, and in the lower panel the reaching of the new stationary regime in $t_f=35\,\mu$s thanks to a proper shaping of the voltage source (see  inset of Fig.~\ref{fig3}b). 

In conclusion, we have shown both theoretically and experimentally the usefulness of reverse engineering to drive at will the current in a RC circuit. It can be easily implemented as a computer-interfacing project.  
From a pedagogical point of view, such studies also contribute to the renewal of the teaching of differential equations with application in the growing field of control in physics.
This method can be readily generalized to other linear circuit such as the RLC circuit. Using the analogy between electricity and classical mechanics, the technique provides interesting and non trivial solutions in this latter domain. For instance, the transport of a particle in a moving harmonic trap obeys the second order linear differential equation:
\begin{equation}
\ddot x + \omega_0^2x=\omega_0x_0^2,
\label{eqqmech}
\end{equation}
where $x$ denotes the position of the particle and $x_0$ that of the bottom ({\it i.e.} the center) of the potential. An optimal transport over a distance $d$ of a particle initially at rest and that reaches its final position at rest imposes the following boundary conditions: $x(0)=0$, $\dot x(0)=0$, $\ddot x(0)=0$, $x(t_f)=d$, $\dot x(t_f)=0$, and $\ddot x(t_f)=0$, with $x_0(0)=0$ and $x_0(t_f)=d$. The position $x(t)$ is chosen by interpolation between the initial and final boundary conditions, and the equation for the motion of the trap is then inferred from Eq.~(\ref{eqqmech}). The method can be further improved to take into account non harmonic traps \cite{PRA2016NL}, or guarantee a robust transport \cite{PRA2014}. Similarly, this idea has been used to drive at will a spin, or two spins to generate for instance entangled states \cite{Qi_2017}.

As presented here, reverse engineering is quite simple and does not require a sophisticated mathematical formalism. 
It is worth emphasizing that it differs from optimal control theory, which aims at extremalizing a
given objective (or cost) function \cite{comfort}, under some constraints \cite{oct,octtocircuits}. Here, the protocols we advocate are
not meant to be optimal, but to perform a given task in a specific, and short, time span. 

Other general methods to speed up quantum transformations have been put forward in the context of quantum mechanics such as the counterdiabatic method  \cite{Counterdiabatic1,Counterdiabatic2}, the Lewis-Riesenfeld invariant methods \cite{LR0,LR},  the fast-forward method \cite{FF} or techniques relying on the Lie algebra \cite{Lie}. Some of those techniques have been recently transposed in the classical world \cite{Jarzynski,PRX} not only in mechanics but also in statistical physics  \cite{Deng,NaturePhys,Li2017}.

\begin{acknowledgments}
It is a pleasure to thank J. Vigu\'{e} and P. Cafarelli for fruitful discussions.  This work was supported by Programme Investissements d'Avenir under the program ANR-11-IDEX-0002-02, reference ANR-10-LABX-0037-NEXT and the LabEx PALM program (ANR-10-LABX-0039-PALM).
\end{acknowledgments}

\end{document}